\documentclass{article}
\usepackage[T1]{fontenc} 
\usepackage[utf8]{inputenc} 
\usepackage{ismir,amsmath,cite,url}
\usepackage{graphicx}
\usepackage{color}

\usepackage[usestackEOL]{stackengine}
\usepackage{subcaption}
\usepackage{enumitem}
\usepackage{kotex}

\usepackage{soul}
\usepackage[usenames,dvipsnames]{xcolor}

\newcommand{\squeeze}{\vspace{-0.1cm}}
\newcommand{\squeezehalf}{\vspace{-0.13cm}}

\newcommand{\MONO}{\texttt{MONO}}
\newcommand{\MIXED}{\texttt{MIXED}}
\newcommand{\CROSS}{\texttt{CROSS}}
\newcommand{\POP}{\texttt{POP}}

\newcommand{\DampVoice}{DAMP-Vocal}
\newcommand{\DAMPMash}{DAMP-Mash}
\newcommand{\MSDSINGER}{MSD-Singer}

\newcommand{\monomono}{\textit{Mono2Mono}}
\newcommand{\mixmix}{\textit{Mix2Mix}}
\newcommand{\monomix}{\textit{Mono2Mix}}

\title{Learning a joint embedding space of \\ monophonic and mixed music signals for singing voice}

\multauthor
{Kyungyun Lee \hspace{1cm} Juhan Nam} {
Graduate School of Culture Technology, KAIST  \\
{\tt \{kyungyun.lee, juhannam\}@kaist.ac.kr}}

\def\authorname{Kyungyun Lee, Juhan Nam}

\begin{document}
\maketitle
\begin{abstract}
Previous approaches in singer identification have used one of monophonic vocal tracks or mixed tracks containing multiple instruments, leaving a semantic gap between these two domains of audio. 
In this paper, we present a system to learn a joint embedding space of monophonic and mixed tracks for singing voice. We use a metric learning method, which ensures that tracks from both domains of the same singer are mapped closer to each other than those of different singers. 
We train the system on a large synthetic dataset generated by music mashup to reflect real-world music recordings. 
Our approach opens up new possibilities for cross-domain tasks, e.g., given a monophonic track of a singer as a query, retrieving mixed tracks sung by the same singer from the database. Also, it requires no additional vocal enhancement steps such as source separation. 
We show the effectiveness of our system for singer identification and query-by-singer in both the in-domain and cross-domain tasks. 



\end{abstract}

\section{Introduction}\label{sec:introduction}
Singing voice is often at the center of attention in popular music. We can easily observe large public interest in singing voice and singers through the popularity of karaoke industry and singing-oriented television shows. 
A recent study also showed that some of the most salient components of music are singers (vocals, voice) and lyrics \cite{demetriou2018vocals}.
Therefore, extracting information relevant to singing voice, i.e., to singers, from music signals, is an important area of research in music information retrieval (MIR) \cite{goto2014singing,humphrey2019introduction}. The relevant tasks include singing voice detection \cite{lee2018revisiting}, singing melody extraction \cite{salamon2014melody,kum2019joint}, singer identification \cite{kim2002singer,mesaros2007singer}, and similarity-based music retrieval \cite{fujihara2005singer,nakano2014vocaltimbre}.

Modern singer information processing systems have been designed to work with only one of monophonic or mixed music signals\cite{mesaros2007singer,lagrange2012robust,wang2018singing}. Then, given both types of signals for analysis, we question whether the system can extract information relevant to singing voice that is transferable between between monophonic and mixed tracks. 
In our experiment, we observe that systems trained with only one type of signals do not perform well, when tested with another type.
To address this limitation, we introduce a new problem of \textit{cross-domain} singer identification (singer-ID) and similarity-based retrieval, in which we regard monophonic and mixed music signals as two different audio domains. 
Cross-domain problems have been explored in computer vision and recommender systems, for example, image retrieval from user sketches to real images\cite{shrivastava2011data} and user preference modeling from movies to books\cite{elkahky2015multi}. In MIR, information transfer between monophonic and mixed tracks can open up new possibilities for singer-based retrieval systems. Some examples are: 
1) given a user's vocal recording in a karaoke application, finding popular singers who sound similar to the user, and 2) given a studio vocal track of a singer, retrieving all tracks (monophonic and mixed) relevant to the singer from a large music database.


To learn a joint feature representation of data from both monophonic and mixed tracks, we adopt a metric learning method, which forces tracks from the same singer to be mapped closer to each other than those from others (\secref{sec:model}). 
To acquire sufficient training data, we create a synthetic dataset by performing a simple music mashup on two public datasets: vocal recordings from DAMP \cite{smith2013correlation} and background tracks from musdb18 \cite{musdb18} (\secref{sec:mashup}). 
We present experiments to demonstrate that our system is able to extract singer-relevant information from both monophonic and mixed music signals, and share the information between the two domains (\secref{sec: Experiments}). 
Source code, trained models, example audios and detailed information about the dataset are available\footnote{http://github.com/kyungyunlee/mono2mixed-singer}. 




\section {Related work} \label{sec:related works}

Cross-domain systems have not yet been examined regarding singing voice analysis. 
Nonetheless, a common challenge in singer information processing systems is to extract singing voice characteristics from music signals in the presence of background accompaniment music.
The most direct way to obtain vocal information is to use monophonic vocal tracks. Recently, Wang et al.\cite{wang2018singing} trained a siamese neural network on monophonic recordings from a subset of the DAMP dataset. Their model scored higher on singer classification but lower on song classification compared to a baseline model. This implies that the model was able to learn singing voice characteristics, rather than the content of a music piece, such as its melody or lyrics. However, since music of our interest is often mixed tracks, this approach has limitations. 

Several works have handled mixed audio signals by enhancing vocal signals through source separation or melody enhancement \cite{fujihara2005singer,mesaros2007singer,lagrange2012robust}. Given recent advances in source separation\cite{stoller2018wave}, this approach may bring improved results for most singing voice analysis systems. 
Another common choice is using audio features that represent human voice or singing voice, such as mel-frequency
cepstral coefficients (MFCCs)\cite{cai2011automatic,lagrange2012robust}. With the success of deep neural networks, it is even possible to learn appropriate features from more general audio representations, i.e., short-time Fourier transform (STFT) or even raw audio. We take this last approach and train our model to be a feature extractor for a given input audio represented by a  mel-spectrogram. 

Depending on the target task, background music can be helpful. An example is singer recognition in popular music\cite{maddage2004singer}. This is because singing style is often dependent on the genre or mood of the music, and singers tend to perform in similar genres throughout their careers. However, our work focuses on learning the actual characteristics of singing voice, independent from background music. 







\vspace{-0.03cm}
\section{Methods} \label{sec:Methods} 
In this section, we describe the data generation pipeline, model configuration and training strategy for learning a joint representation of monophonic and mixed tracks for singing voice. 

\subsection {Data generation} \label{sec:data generation} 
For training cross-domain singer-ID and retrieval systems, a sufficiently large number of monophonic and mixed track pairs per singer is needed.
Existing singing voice datasets, such as MIR-1K\cite{hsu2010improvement}, iKala\cite{chan2015vocal} and Kara1K\cite{bayle2017kara1k}, provide the monophonic and mixed track pairs, but they have a small number of singers or only a few tracks per singer. An alternative option may be to perform singing voice detection (SVD) and vocal source separation on a large dataset, but the audio quality can be degraded. 
 
In this work, we choose to utilize the DAMP dataset, which contains vocal-only recordings from mobile phones of around 3,500 users from the Smule karaoke app (there are 10 full-length songs per user). This serves as the main ingredient to generate our synthetic singer dataset. As a preprocessing step, we perform a simple energy-based SVD to remove silent segments. Then, 1000 singers are chosen for training stage and additional 300 singers are put aside for testing. The original DAMP dataset processed with SVD, \textit{\DampVoice{}}, is used as the monophonic dataset; the synthesized mixed track dataset, \textit{\DAMPMash{}} (detailed in section \ref{sec:mashup}), is used as the mixed track dataset in this work.

\subsubsection{Mashup: DAMP and musdb18}\label{sec:mashup}
\begin{figure}[t]
\centering
 \includegraphics[width=\columnwidth]{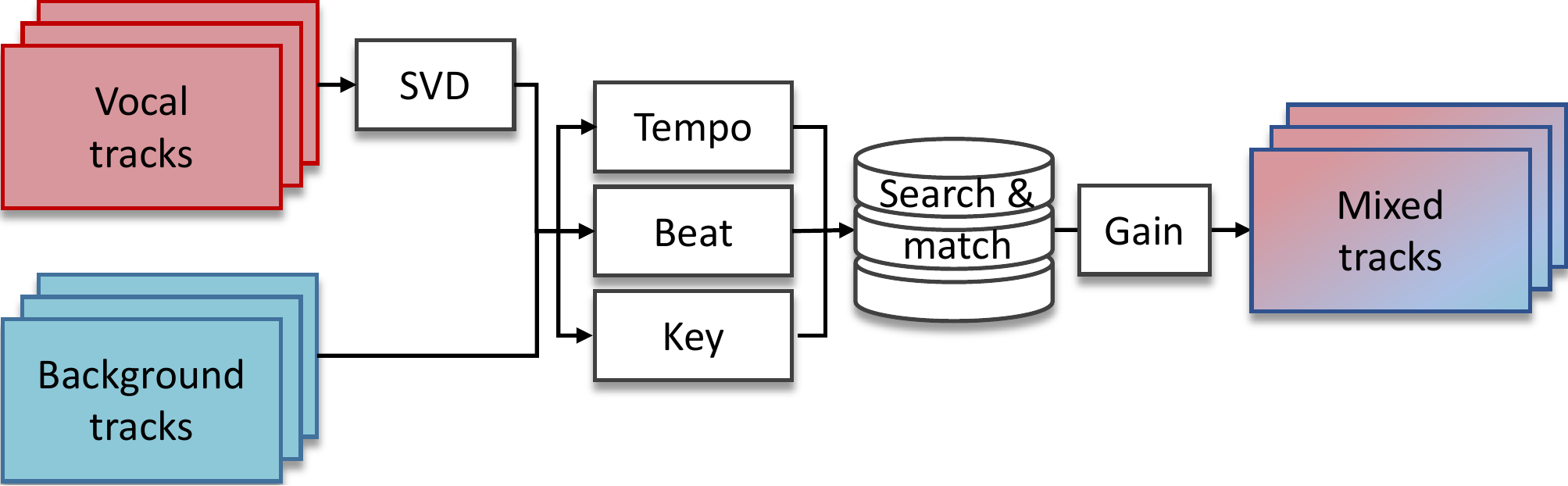}
  \caption{Mashup pipeline to generate synthetic dataset, \textit{DAMP-Mash}.}
  \label{fig:mix pipline}
\end{figure}

A music mashup is a way of creating music by carefully mixing two or more tracks from several different pre-recorded songs. 
Inspired by such work, we automatically generate a synthetic singer dataset, called \DAMPMash{}, by combining vocal recordings from the DAMP dataset with background tracks from the musdb18 dataset. 
Instead of random mixing, we build a pipeline (\figref{fig:mix pipline}) to identify the "mashability"\cite{davies2014automashupper} between tracks. 
Our mashability criteria requires 3-second long vocal and background tracks to have the same tempo and key. 
Once the two segments pass the mashability test, they are mixed at their nearest beat location. Before mixing, we adjust the loudness by balancing the root-mean-square energy between both segments. 

Tempo detection and beat tracking are performed at track-level using librosa 0.6.2 \cite{brian_mcfee_2018_1342708}. On the other hand, key is determined locally at 3-second long segments by using the Krumhansl-Schmuckler key finding algorithm\cite{professor1990cognitive} on chromagrams. As a result, vocal segments within the same song end up being mixed with multiple different background tracks. Thus, we view our synthetic dataset as being genre-independent. This mashup pipeline can be also regarded as a  data augmentation technique. 

\subsection{Model} \label{sec:model}

\subsubsection {Skeleton model}\label{sec:skeleton} 
A 5-layer 1-D convolutional neural network (CNN) is the skeleton of larger networks used in this paper. First four convolutional layers have 128 filters of size 3, each followed by a maxpooling layer of size 3. The last convolutional layer consists of 256 filters of size 1 to output a final embedding vector of 256 dimensions. All convolution operations are done on the temporal dimension only. The final embedding vector will be used to represent input audio and to perform tasks described in \secref{sec: Experiments}. Batch normalization and Leaky ReLU\cite{maas2013rectifier} are applied to all convolutional layers, and dropout of 50\% is applied after the last convolutional layer to prevent overfitting. 

The input is a 3-second long audio of at least 70\% singing voice frames. The sample rate of audio is 22050 Hz and we compute an STFT using a  1024-sample long Hanning window with 50\% overlap.  We then convert it to a mel-spectrogram with 128 bins and apply logarithmic compression on the magnitude. As a result, the input shape is 129 frames by 128 bins.



\subsubsection {Embedding model} \label{sec:siamese model} 
\begin{figure}[t]
\centering
 \includegraphics[width=1.0\columnwidth]{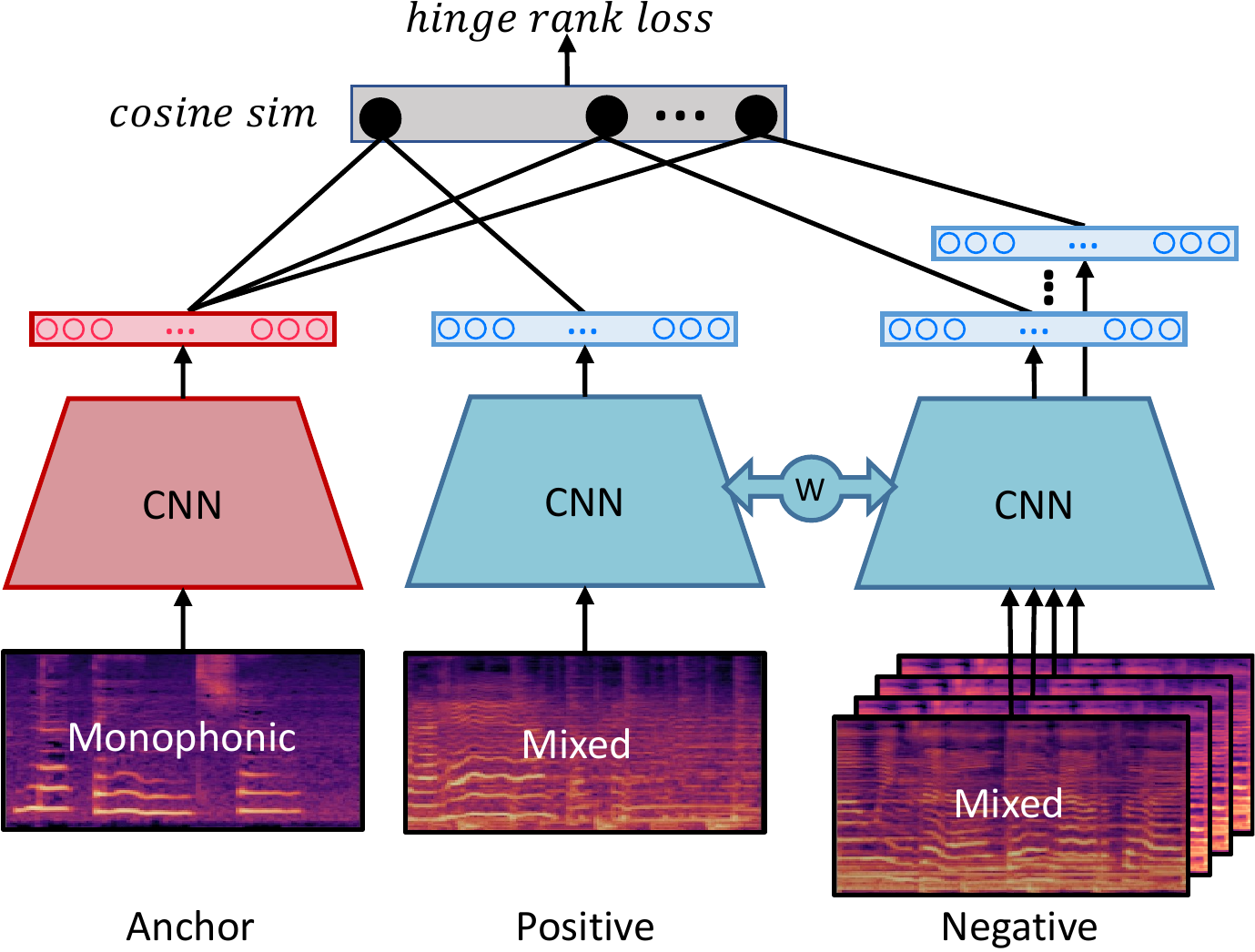}
  \caption{Configuration of \CROSS{} model. The anchor network (left) is for modeling monophonic tracks; the rest is for modeling mixed tracks.}
  \label{fig:model}
\end{figure}

The outcome of metric learning is a mapping function from inputs to output vectors in an embedding space, where inputs of the same class are closer to each other than those from different classes. 
Specifically, we build our model upon a triplet network, 
which consists of three potentially weight-sharing networks that takes three inputs: anchor, positive (same class as the anchor) and negative (different class as the anchor) items. This architecture can be extended to take multiple negative items \cite{sohn2016improved} to overcome the limitation of learning from only one negative example. 
For model configuration, we closely follow the work of \cite{park2017representation}, using 4 negative items. We also use a type of margin-based ranking loss, called hinge rank loss, with cosine similarity as our metric \cite{frome2013devise}.

Thus, the loss function for a given set of anchor ($p_i$), positive ($p_{+})$ and negative ($p_{-}$) feature vectors is: 
\begin{equation} 
\squeeze
loss(p_i,p) = \sum_{p_{-}}max[0, \alpha - S(p_i,p_{+}) + S(p_i,p_{-})]
\squeeze
\end{equation} 
where $S$ is a similarity score: 
\begin{equation}  \label{eq:cosine}
\squeeze
S(p_i, p_j) = cos(p_i, p_j) = \frac { p_i \cdot  p_j}{||p_i|| \cdot ||p_j||}
\end{equation} 
and 
$\alpha$ indicates the margin, which is fixed to 0.1 after performing a grid search on values between 0.01 and 1.0.  
Negative tracks are selected through negative sampling among tracks that do not belong to the singer of the anchor item. We tested a more difficult negative sampling strategy of selecting the four highest scoring items among twenty randomly chosen negative samples, but the model showed minor improvement with an increase in computation time. Investigation on negative sampling is left as our future work.

We choose metric learning for three main reasons. First, by giving a higher similarity score to any pair of tracks performed by the same singer, the model can learn to identify the singer from a track regardless of it being monophonic or mixed. Thus, it is especially suitable for training cross-domain systems. Second, using singer identity as the only ground truth to measure similarity between two tracks will force the model to focus only on singing voice. Since DAMP-Mash is genre-invariant, the only common component in two tracks is going to be related to singing voice. Thus, we may expect the model to perform a feature-level source separation on mixed tracks. Lastly, the model can be trained on a larger number of singers without increasing the number of parameters. On the other hand, a classification model that uses a softmax layer will need to increase the output layer size to match the number of training singers \cite{park2017representation}. 

We explore our ideas with three models, which differ in the type of data used for training: 
\squeezehalf
\begin{itemize}[leftmargin=*]
\item\MONO{}: all inputs are monophonic tracks
\squeezehalf
\item\MIXED{}: all inputs are mixed tracks
\squeezehalf
\item\CROSS{}: anchors are monophonic, while positive and negative items are mixed tracks (Figure \ref{fig:model})
\end{itemize}
\squeezehalf
Our main idea in this work is reflected in the \CROSS{} model, for which the hinge rank loss ensures that the cosine similarity between monophonic and mixed tracks from the same singer is scored higher than that from different singers. \MONO{} and \MIXED{} models are reference models for comparison.

While networks within \MONO{} and \MIXED{} model share weights, in \CROSS{} model, the anchor network and the rest do not share weights. Thus, it yields two separate feed-forward networks, each designed specifically for its corresponding domain (\figref{fig:model}). As a result, depending on the domain of an input audio, one of the two networks is used as the feature extractor. Each network is configured with the skeleton model described in \secref{sec:skeleton}. 

\subsubsection{Pre-training via classification}\label{sec:pretrain}
Metric learning is known for its difficulty in optimization\cite{wang2017deep,zhang2016embedding}. To alleviate this problem, we train a classification model and use it to initiate the learning of the embedding models. The classification model has one linear layer added to the skeleton model (\secref{sec:skeleton}) and predicts the correct singer with a softmax probability. Instead of fully training it, we remove the last output layer after 30 epochs and use it to continue the training in a metric learning style. We do not freeze any layers.

\section{Experiments \& Evaluation} \label{sec: Experiments}

\subsection{Test scenarios}\label{sec:task setup}
Two main tasks for evaluation are singer identification and query-by-singer. 
In both tasks, a music signal to be analyzed (source) is queried to a collection of data (target) to retrieve desired information. Depending on the domain of source and target data, we design three test scenarios: 
\begin{itemize}[leftmargin=*] \squeeze
\item \monomono: both source and target data are \textit{monophonic} (in-domain)
\squeeze
\item  \mixmix:  both source and target data are \textit{mixed} (in-domain)
\squeeze
\item \monomix:  source data is \textit{monophonic}, but the target data is \textit{mixed} (cross-domain)
\end{itemize}
\squeeze 

Each task is evaluated on all three test scenarios. 

\subsection{Task 1: Singer identification} \label{sec:singer rec}

\textbf{Dataset} : 
We select 300 singers unseen from the training stage for evaluation.
For each singer, we use 6 tracks for building singer models and set aside 4 tracks as query tracks, resulting in 1200 queries. Depending on the domain of source and target data, \DampVoice{} (monophonic) and \DAMPMash{} (mixed) dataset are used accordingly. 

\hspace{1pt}

\noindent\textbf{Description} :
As in \cite{Park2018AHO, royo2018disambiguating}, singer identification is to determine the correct singer of the query track among the 300 candidate singer models. All queries and singer models are represented as 256 dimensional feature vectors; a track vector is an average of 20 feature vectors computed from 3-second long segment of the same track and a singer model is an average of 6 track vectors from the same singer. We made predictions by computing cosine similarity \eqref{eq:cosine} between the query track and all the singer models. Then, the singer with the highest score is chosen. 

For our baseline, we train a Gaussian Mixture Model-Universal Background Model (GMM-UBM), which is commonly used in speaker recognition systems \cite{reynolds2000speaker}. Each singer model is adapted through maximum a posteriori (MAP) estimation from a single singer-independent background model. All models are composed of 256 components with MFCCs of dimension 13 as input. We train 2 GMM-UBMs, one with monophonic tracks for \monomono{} and the other with mixed tracks for \mixmix{}.

We report both top-1 and top-5 classification accuracy. They represent the proportion of correct guesses out of 1200 queries in total. Top-5 accuracy is calculated by considering a prediction as being correct if the ground truth singer appears within the top 5 highest scoring singers.

\subsection{Task 2: Query-by-singer}\label{sec:retrieval}
\textbf{Dataset} : As in the singer recognition task (\secref{sec:singer rec}), same 300 singers are used for evaluation. 6 tracks from each singer are selected to build a collection of 1800 tracks to represent a search database and 4 tracks are used as test queries. 

\hspace{1pt}

\noindent\textbf{Description} :
Given a query track, the task is to retrieve tracks that are performed by the same singer among the track database. We compute the similarity (Equation \eqref{eq:cosine}) between the query track and all the tracks in the database, and rank them based on their similarity scores. 
This can be applied to singer-based music recommender systems to discover singers with similar singing voice characteristics. 

We report \textit{precision} and \textit{recall-at-k} as well as mean average precision (mAP) score, where \textit{k} is set to 5 to resemble music retrieval systems. 
Given a query track performed by singer \textit{A}, \textit{precision-at-k} (Pr@k) refers to the proportion of tracks that are performed by \textit{A} and are recommended among \textit{k} items; \textit{recall-at-k} (R@k) refers to the proportion of tracks that are performed by \textit{A} and are recommended out of all the tracks performed by \textit{A} (6 tracks in total) in the database. Unlike Pr@k and R@k, mAP takes into account the actual order of the recommended tracks. Thus, it is useful for music recommender systems, where it is important that relevant items are not only retrieved, but also with higher confidence than false positive items. 


\subsection{Results}

\begin{figure}[t]
\centering
 \includegraphics[width=\columnwidth]{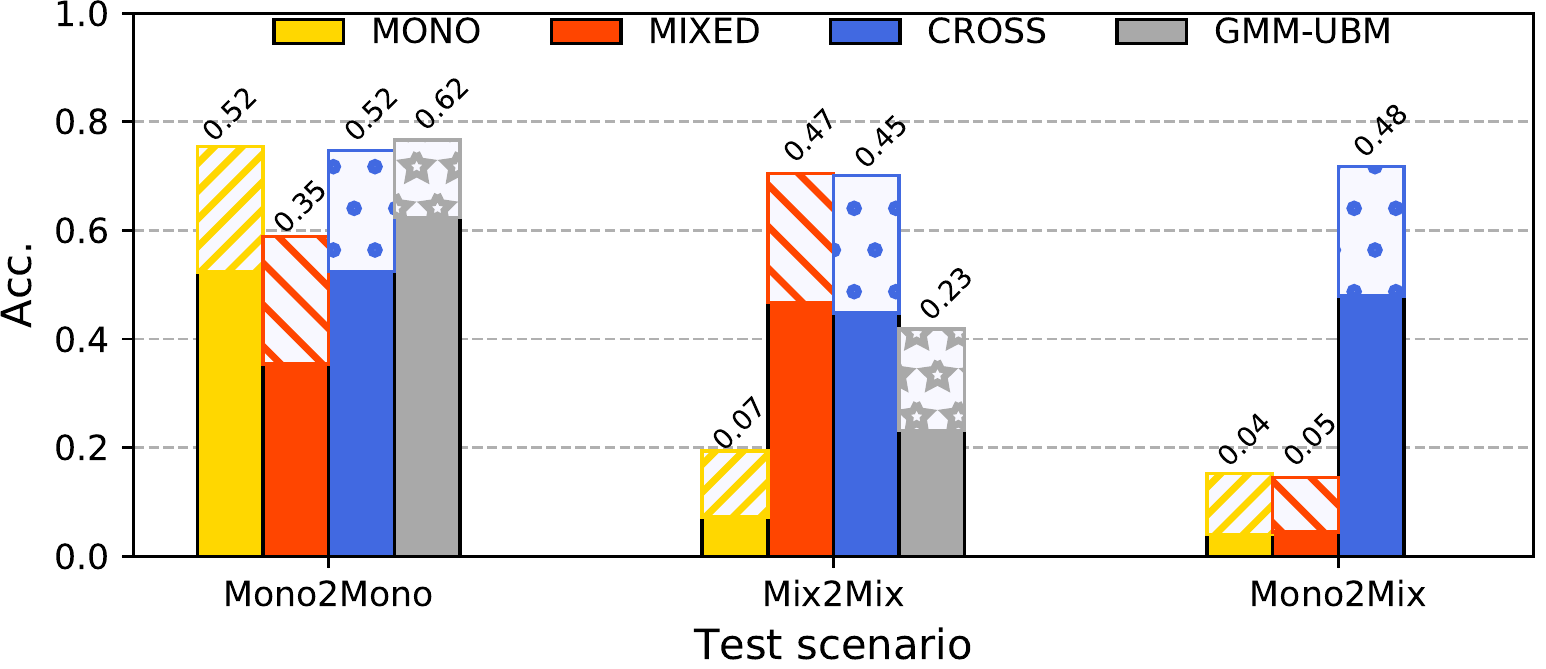}
  \caption{Singer identification results for \MONO{}, \MIXED{} and \CROSS{} on different test scenarios. The solid section points to the top-1 accuracy (also written above each bar) and the hatched section points to the top-5 accuracy.}
  \label{fig:singer id}
\end{figure}

\begin{figure*}[t]
  \begin{subfigure}[b]{1\linewidth}
    \includegraphics[width=\columnwidth]{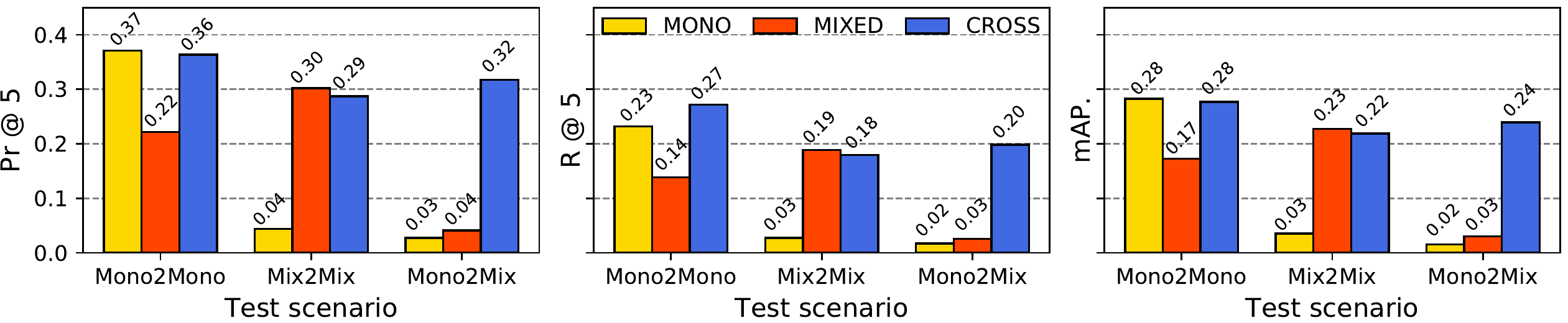}
  \end{subfigure}
  \caption{Query-by-singer result for \MONO{}, \MIXED{} and \CROSS{} models. \textit{precision-at-k} (left), \textit{recall-at-k} (center) and mean average precision score(right) are shown. Each number above the bar refers to corresponding model's score.}
  \squeeze
  \label{fig:precision}
\end{figure*}

\begin{figure}[t]
\centering
 \includegraphics[width=\columnwidth]{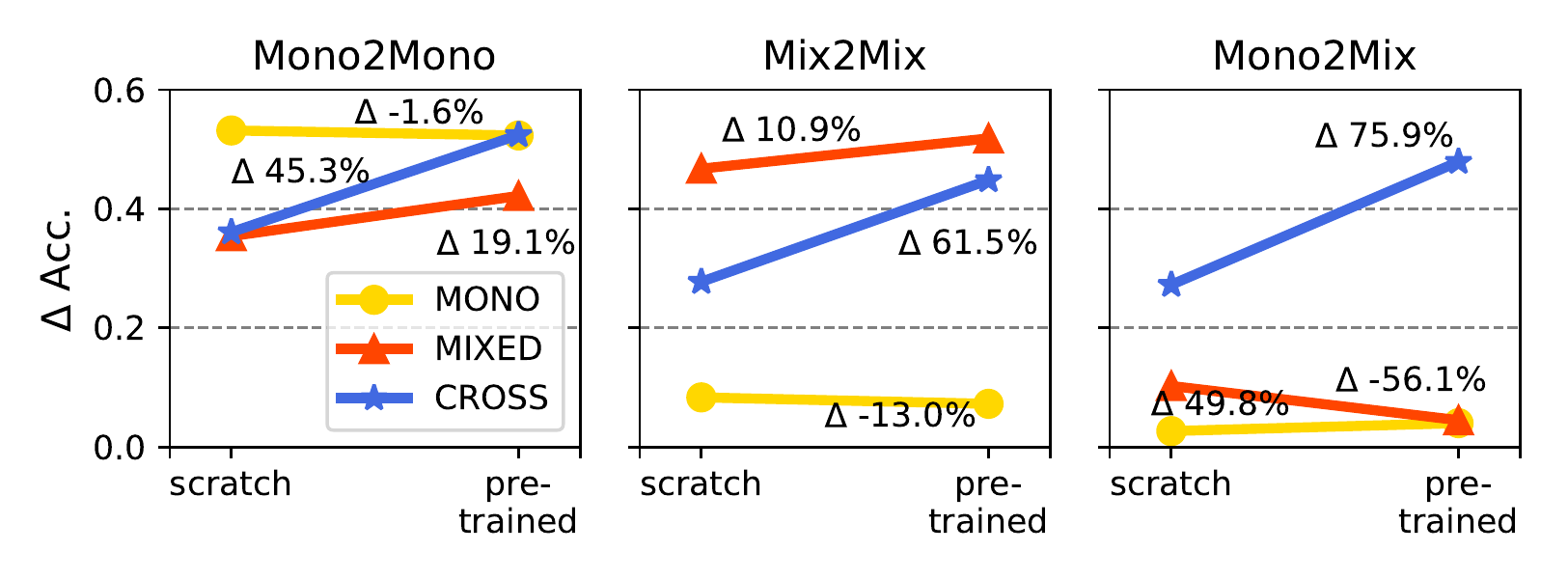}
  \caption{Top-1 accuracy improvement on singer identification when using a pre-trained network. The numbers on the line indicates the percentage of improvement.}
  \label{fig:pretraining}
\end{figure}
\squeezehalf

\begin{figure}[t]
\centering
 \includegraphics[width=\columnwidth]{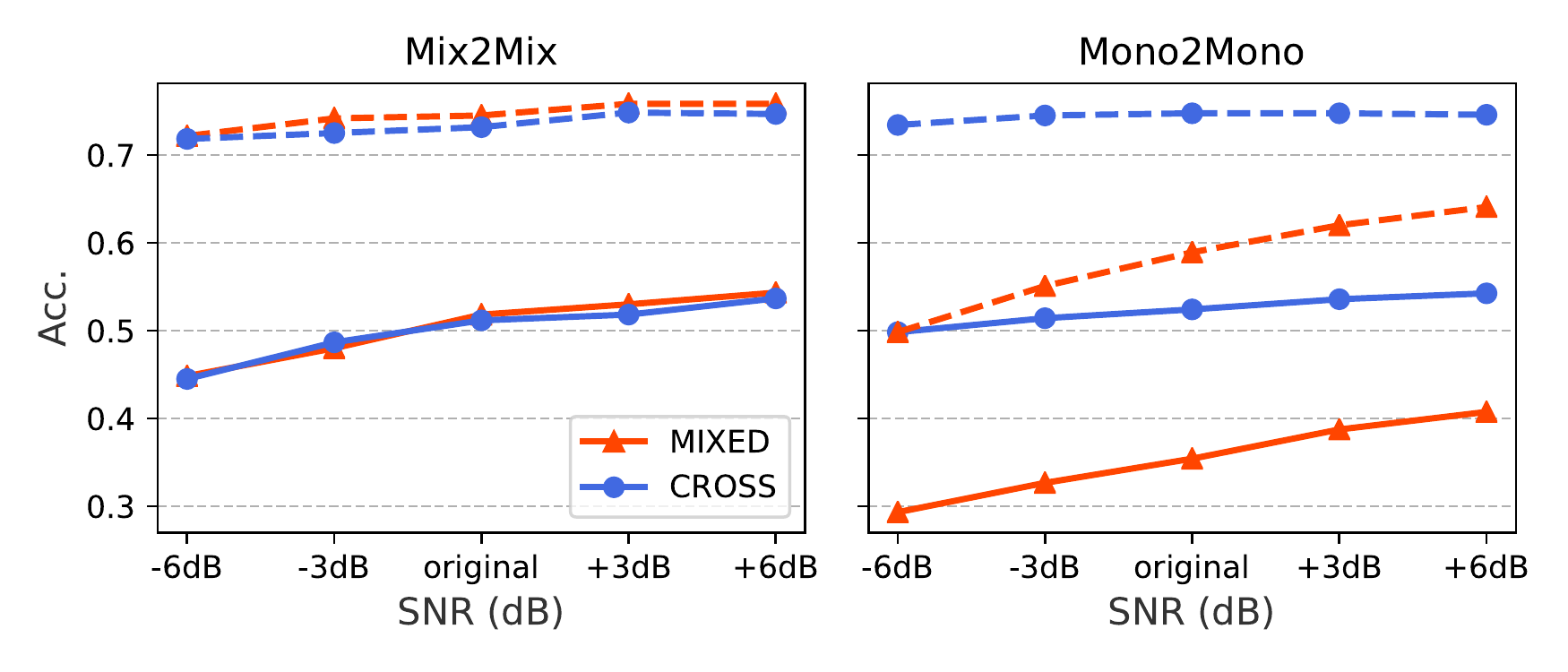}
\caption{Result of \MIXED{} and \CROSS{} model for singer recognition on varying SNR. Solid line indicates top-1 and dotted line indicates top-5 accuracy.}
  \label{fig:snr}
\end{figure}

In \figref{fig:singer id} and \figref{fig:precision}, we observe a large performance variation for \MONO{} and \MIXED{} models across different test scenarios. Both models perform well in \monomono{} and \mixmix{}, respectively. However, their performances drop significantly in other scenarios, especially for \monomix{}. 
This is expected, because these models have not been trained to handle cross-domain scenarios. 

On the other hand, \CROSS{} model performs well on all three test scenarios, benefiting from two jointly trained networks that can each handle monophonic and mixed tracks. We see that it is the only model that is able to match and compare the singer identity between tracks from different domains. Also, its performances on \monomono{} and \mixmix{} are on par with the \MONO{} and \MIXED{} models. This is a useful observation, since training only the \CROSS{} model can still give good performance on all three test scenarios, avoiding the effort of training separate models for each scenario. 
Note that the baseline model, GMM-UBM, shows the best performance in \monomono{}, but not so well in \mixmix{}. Result for \monomix{} is omitted, since it is close to random prediction. When there is no background music, GMM-UBM with MFCCs are efficient in characterizing singing voice. 

As mentioned in \secref{sec:pretrain}, we show the effect of using a pre-trained network on singer identification task (\figref{fig:pretraining}). \CROSS{} model (blue stars) shows the largest performance improvement compared to the other two models. We assume that comparing the singer identity between monophonic and mixed track is more difficult than comparing between tracks of the same domain. Therefore, a pre-trained model, which learned to somewhat identify singers from an input audio, serves as a hint to focus on signals relevant to singing voice. Using a pre-trained model not only improved the accuracy, but also accelerated the learning process. 

Regarding background music as noise and singing voice as the signal, signal-to-noise ratio (SNR) has a large impact on the performance of singing voice analysis systems\cite{lee2018revisiting}. 
We change the SNR of the test data and show results on singer recognition task for \MIXED{} and \CROSS{} models in \figref{fig:snr}. 
Since \monomono{} deals with only monophonic tracks, the change in performance exhibited in \monomono{} (right) is due to the overall loudness of the track, not SNR. Therefore, as the performance change on \mixmix{} (left) shows a similar trend across different SNR, it implies that models trained on DAMP-Mash dataset is able to identify singing voice in more difficult conditions. This is a great benefit for singing voice analysis systems.


\subsection {Evaluation on Popular Music Recordings}

As our system is trained with a synthetic dataset, we evaluate it on popular music recordings to ensure that the trained system can also generalize to real-world data.

\hspace{1pt}

\noindent\textbf{Dataset} : 
Million Song Datatset (MSD) contains 1,000,000 tracks and 44,745 artists from popular music recordings. As done in \cite{Park2018AHO}, we filter the dataset to select artists with substantial vocal tracks using singing voice detection (SVD). This dataset is referred to as \textit{MSD-Singer}\footnote{Details provided at http://github.com/kyungyunlee/MSD-singer}.  
For comparison, we train a model, named \texttt{POP}, on 1000 artists from MSD-Singer dataset. We used 17 30-second long tracks for each artist for training.
500 singers, unseen from the training stage, are used for evaluation. 15 tracks of each singer are used for building singer models and 5 tracks are used as query tracks. 

\hspace{1pt}

\noindent\textbf{Description} : 
The task is equivalent to singer identification in Section \ref{sec:singer rec}, only with a different dataset. The result from the  \texttt{POP} model is the upper bound, as it is trained and tested on MSD-Singer dataset; meanwhile, \MIXED{} and \CROSS{} models are trained on \DAMPMash{} and \DampVoice{} dataset. 

\hspace{1pt}

\noindent\textbf{Result}: 
The result shown in Table \ref{table:pop_result} compares \MIXED{} and \CROSS{} models with \POP{} model and a random classifier. It shows that even though our models are trained only with the synthetic dataset, they are also able to identify singing voice in popular music. Therefore, we can confirm that DAMP-Mash dataset is able to represent the popular music to some degree and that our models are able to generalize to real-world recordings. We believe that the results will improve with a better automatic mashup pipeline. Training the \CROSS{} model on source separated MSD-Singer dataset is also left as a future work.

\begin{table}[]
\begin{tabular}{c|cc|c|c}
          & \MIXED    & \CROSS    & \POP  &  Random\\ \hline
Acc.       & 0.291 & 0.282 & 0.393 & 0.002  \\
Top-5 Acc. & 0.511 & 0.491 & 0.664 & 0.01
\end{tabular}
\caption{Accuracy result on singer recognition on dataset of popular music recording, \MSDSINGER dataset}
\label{table:pop_result}
\end{table}\label{sec:pop music}

\begin{figure}[t]
\centering
 \includegraphics[width=\columnwidth]{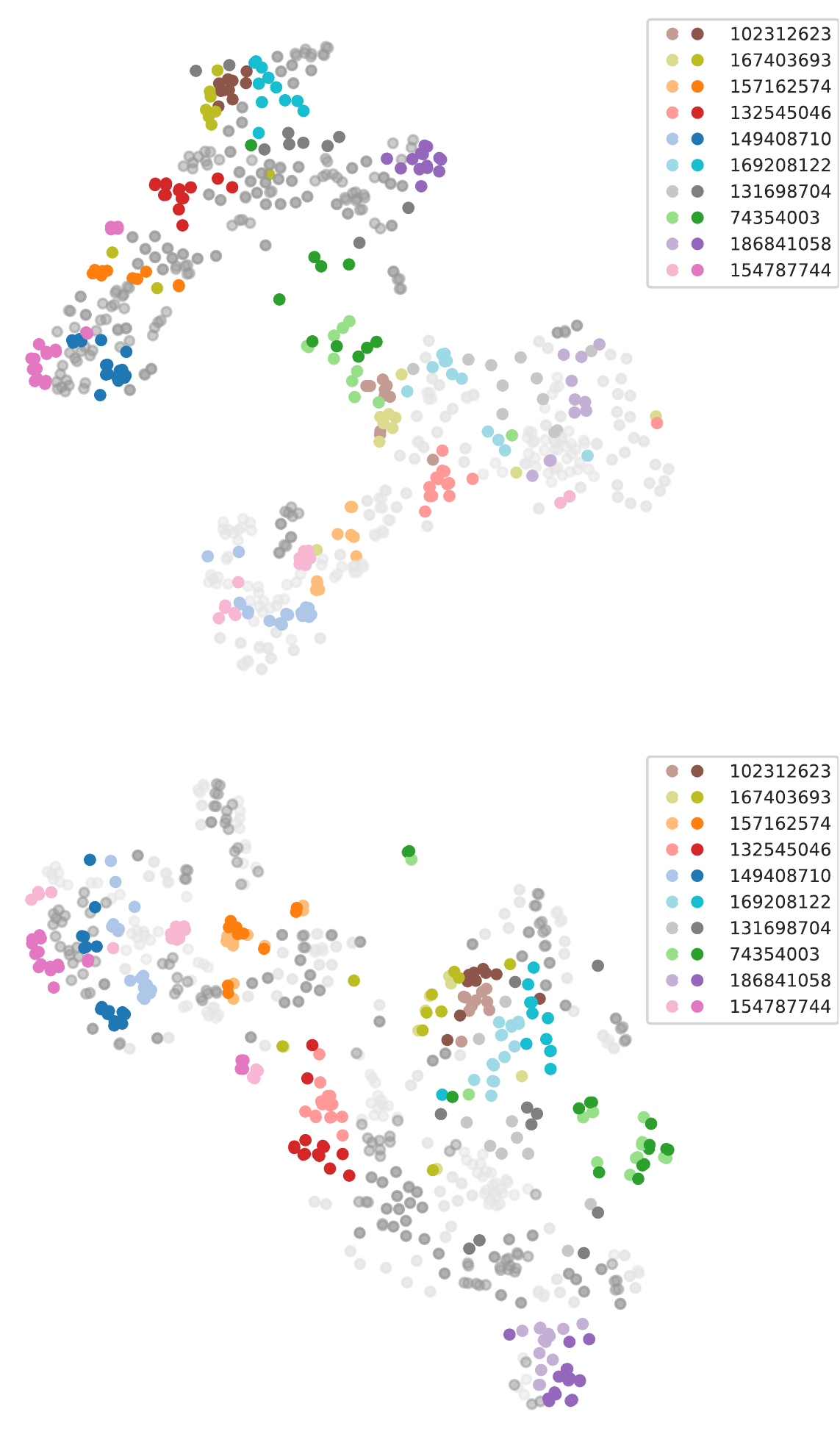}
  \caption{Singer embedding space from \MIXED{} model (top) and \CROSS{} model (bottom). The label numbers are player IDs from the DAMP dataset. The colors on the left column refers to monophonic vocal tracks; the right column refers to mixed tracks. Best viewed in color.}
  \label{fig:mixed_cross_tsne}
\end{figure}


\section{Embedding space visualization} 

We visualize the embedding space learned by the \MIXED{} and \CROSS{} models to understand how they each process monophonic and mixed tracks. From DAMP-Voice and DAMP-Mash dataset, we select 25 singers unseen from the training stage and highlight 10 with colors for better visualization. 20 tracks are plotted for each singer: 10 monophonic vocal tracks and their corresponding mixed tracks. After feature extraction, we reduce the dimension of each feature vector from 256 to 2 dimensions using t-distributed stochastic neighbor embedding (t-SNE) \cite{maaten2008visualizing}. Each dot on the embedding space represents a track. For visualization, we use a paired color palette and assign lighter color to monophonic tracks. Since both monophonic and mixed tracks are from the same singer, an ideal embedding space will show clusters of 20 tracks for each singer. 

\figref{fig:mixed_cross_tsne} shows the embedding space learned from the \MIXED{} model. There is a noticeable gap between the features of monophonic tracks and that of mixed tracks, which means that the model differentiates monophonic and mixed tracks, rather than finding similar singing voice. Still, we can see that the model is able to cluster tracks from the same singer within the same domain.
However, in \figref{fig:mixed_cross_tsne}, the monophonic and mixed track features of the same singer are mapped close to each other. This explains why the \CROSS{} model shows good performance on cross-domain tasks. We can observe that it is able to transfer singer information across two domains.


\section{Motivation for future work}\label{sec:directions}


\noindent\textbf{Improvement on music mashup}: Our mashup pipeline has a large room for improvement. Besides errors produced from existing algorithms, such as key detection, more efforts can be put towards mixing two tracks with a good balance as in real-world recordings. 
A good automatic mashup system can benefit many areas of research in MIR. The creativity and limitless choices of techniques that can be applied to generate a mashup imply that a large amount of multitrack dataset can be generated for many tasks of interest. 

\hspace{1pt}

\noindent\textbf{From track to singer modeling}: In this work, we use an average of several track-level feature vectors to build singer models. However, in case of singers with highly varying vocal characteristic between different tracks and taking into account the ``album effect'', averaging may not always be the best choice. Exploring GMMs with multiple mixtures or principal component analysis (PCA) can be an interesting future direction. 

\hspace{1pt}

\noindent\textbf{Going beyond singing voice}: Although we have focused on singing voice, our methods can be tested with tasks involving other instruments, such as multiple instrument recognition. The same mashup technique can be applied to create a dataset, by replacing the monophonic vocal tracks with any instrument of interest. Data generation with mashup may yield better results for instrument recognition in real-world recordings compared to the method proposed in \cite{tokozume2017learning}, where only two monophonic instrument tracks are used to create a random mix. 

\section{Conclusion}
In this paper, we introduced a new problem of cross-domain singer identification and singer-based music retrieval to allow information transfer between monophonic and mixed tracks. 
Through data generation using music mashup, we were able to train an embedding model to output a joint representation for singing voice from tracks regardless of their domain. We evaluated on three different test scenarios, which include both in-domain and cross-domain cases. A huge advantage of \CROSS{} model is that it performs well not only on the cross-domain scenario, but also on commonly observed in-domain scenarios. Therefore, by training only the \CROSS{} model, it yields two models, one for each domain. Additional evaluation on varying SNR and on popular music dataset demonstrated that the model is robust to background music and can also be generalized beyond our synthetic dataset. 

To conclude, we believe that cross-domain systems can enable many interesting applications related to singing voice, as well as in MIR. Specifically, our future interests include improving the quality of the mashup dataset and performing comparisons between singing voices of karaoke users and that of popular singers. 

\vspace{5mm} 

\section{Acknowledgements}
We thank Keunwoo Choi for valuable comments and reviews. This work was supported by Basic Science Research Program through the National Research Foundation of Korea funded by the Ministry of Science, ICT \& Future Planning (2015R1C1A1A02036962), and by NAVER Corp.

\bibliography{ISMIRtemplate}

%
%
%
%

\end{document}